# Adiabaticity violation under arbitrarily slow evolution

Oubo You[1,2]*†, Zhaoqi Jiang[1,3,4]†, Jinhui Shi[1,3,4], Qing Dai[2], Chunying Guan[1,3,4]* & Shuang Zhang[1,5,6]*

[1]New Cornerstone Science Laboratory, Department of Physics, University of Hong Kong, Hong Kong, 999077, China.

[2]School of Materials Science and Engineering, Shanghai Jiao Tong University, Shanghai, 200240, China

[3]Key Laboratory of In-Fiber Integrated Optics of Ministry of Education, College of Physics and Optoelectronic Engineering, Harbin Engineering University, Harbin, 150001 China

[4]Key Laboratory of Photonic Materials and Devices Physics for Oceanic Applications, Ministry of Industry and Information Technology of China, College of Physics and Optoelectronic Engineering, Harbin Engineering University, Harbin, 150001 China

[5]Department of Electrical and Electronic Engineering, University of Hong Kong, Hong Kong, 999077, China.

[6]Materials Innovation Institute for Life Sciences and Energy (MILES), HKU-SIRI, Shenzhen, P.R. China

*Corresponding author. Email: yououbo@sjtu.edu.cn (O. Y.); guanchunying@hrbeu.edu.cn (C. G.); shuzhang@hku.hk (S. Z.)

†These authors contributed equally to this work.

## Abstract

The quantum adiabatic theorem, a cornerstone of quantum mechanics, asserts that a gapped quantum system remains in its instantaneous eigenstate during sufficiently slow evolution, provided no resonances occur. Here we challenge this principle and show that adiabaticity can be violated even in arbitrarily slow processes. We introduce two new parameters, Instantaneous Transition Accumulation (ITA) and Instantaneous Transition Probability (ITP), to redefine the framework of adiabatic evolution. These parameters, grounded in cross-Berry connections and eigenstate amplitudes, reveal the dynamic and geometric factors governing adiabaticity. Using a new Phase Difference Manipulation (PDM) method, we control ITP and ITA to induce adiabaticity violation in a Landau-Zener (LZ) process. We experimentally demonstrate this counterintuitive phenomenon in a photonic waveguide system, where a slow LZ process defies adiabaticity, switching energy levels despite a fivefold slower evolution speed than a conventional adiabatic process. This discovery reshapes our understanding of quantum evolution and holds potential for quantum computing, topological physics, and photonic technologies.

# Main

**Introduction**

The quantum adiabatic theorem, established by Max Born and Vladimir Fock in 1928, stands as a cornerstone of quantum mechanics, governing the behavior of gapped quantum systems under slow parameter changes [1,2]. From its origins in early quantum theory to its modern applications in adiabatic quantum algorithms and topological phases, the theorem asserts that a system remains in its instantaneous eigenstate if its Hamiltonian evolves sufficiently slowly and avoids resonances [3-8]. This principle has catalyzed transformative technologies and concepts, including adiabatic quantum computation [9-13], where ground states encode solutions to optimization problems; topological insulators [14-18], where adiabatic evolution preserves Hall effect; and quantum state preparation, such as Bose-Einstein condensate formation [19-21]. In both quantum and classical systems, adiabatic processes underpin topological pumps [22-31], enabling robust energy manipulation in waveguide systems or mode transformation. In recent years, a technique called shortcut to adiabaticity [32-40] has been developed to increase the robust evolution speed.

Here, we uncover a surprising exception to this assumption in quantum mechanics: adiabaticity can fail even in arbitrarily slow, resonance-free processes. This counter-intuitive discovery challenges a century-long tenet of quantum mechanics, revealing that the traditional adiabatic criterion, based on the ratio of cross-Berry connections (geometric phase factors) to energy gaps, is incomplete. To address this gap, we introduce two revolutionary parameters: Instantaneous Transition Accumulation (ITA) and Instantaneous Transition Probability (ITP). These parameters redefine adiabaticity by capturing the interplay of dynamic eigenstate amplitudes and geometric properties, such as Berry connections and quantum metrics, in evolving quantum systems. Unlike the conventional framework, ITA and ITP provide a comprehensive description of population transfers between eigenstates, offering a new lens for understanding and controlling quantum evolution.

To manipulate adiabaticity with unprecedented precision, we develop a new Phase Difference Manipulation (PDM) method. By strategically pausing the Hamiltonian's evolution, PDM controls ITA without altering the parameter path, enabling deliberate violations of adiabaticity. We test this framework in a spoof surface plasmon polariton (sSPP) waveguide system, a versatile platform for probing quantum phenomena with classical light. Our experiments demonstrate adiabaticity violation in a Landau-Zener (LZ) process [41], where the system switches energy levels despite an evolution speed five times slower than a standard adiabatic process. This finding not only reshapes our understanding of quantum dynamics but also carries profound implications for adiabatic quantum computation, where ground-state fidelity is critical; topological physics, where slow variations may disrupt robust states; and photonic technologies, where waveguide systems enable scalable applications.

**Breaking adiabaticity in slow LZ process**

We begin by discussing the standard adiabaticity condition for a typical LZ process, which describes the transition probability between two quantum states during time evolution. Without losing generality, we consider a two-level system with the Hamiltonian $H_{LZ}(t) = V\sigma_x + \lambda\sigma_z$, where $\sigma_{x,z}$ are Pauli matrices, $\lambda(t) = \alpha t + \beta$, $V$ is the coupling strength and $\alpha$, $\beta$, $V$ are constants. The instantaneous eigenenergies are $E_\pm(\lambda) = \pm\sqrt{V^2 + \lambda^2}$ with corresponding eigenstates $|\phi_\pm(\lambda)\rangle = N_\pm(\lambda)(\lambda + E_\pm, V)^T$, where $N_\pm(\lambda)$ are normalization constants. Ideally, the LZ process starts from one of the eigenstates at $\lambda = -\infty$, but this is impractical for numerical and experimental studies. Instead, we set $\lambda(t = 0) = -1.5$ and ends at $\lambda = 1.5$, shown in Fig. 1a. The time-dependent evolution of an initial state $|\Phi_0\rangle = |\Phi(0)\rangle$ can be written as $|\Phi(t)\rangle = U(t)|\Phi_0\rangle$, where $U(t) = \mathcal{T}exp\left(\int_0^t -iH_{LZ}(\lambda(t))\,dt\right)$ is the time evolution operator and $\mathcal{T}$ denotes time-ordering. At each time point $t$, the evolved state can be decomposed into the instantaneous eigenstates, i.e. $|\Phi(t)\rangle =$

$c_-|\phi_-(\lambda)\rangle + c_+|\phi_+(\lambda)\rangle$. For a standard LZ process shown by the blue curved arrow line in Fig. 1a, the initial state is $|\Phi_0\rangle = |1\rangle = (0,1)^T \approx |\phi_+(-1.5)\rangle$, evolving at a speed $\lambda' = \alpha = d\lambda/dt = 0.2$. The variation of $\lambda$ with respect to time $t$ is shown by the blue line in Fig. 1b. Throughout the process, $|\Phi(t)\rangle$ remains approximately aligned with $|\phi_+(\lambda)\rangle$ indicating adiabaticity at sufficiently low speed.

Next, we demonstrate how adiabaticity can be violated at arbitrarily slow evolution rates by modifying the evolution process. Specifically, we consider an evolution where $\lambda(t)$ pauses at selected points along the path for appropriate time intervals, shown by the yellow and red broken line in Fig. 1b. Remarkably, the evolved state $|\Phi(t)\rangle$ no longer tracks $|\phi_+\rangle$ but transitions to $|\phi_-\rangle$, as shown by the yellow line with red points in Fig. 1a. In this modified process, the maximum instantaneous speed, given by the slopes of the yellow segments in Fig. 1b, remains 0.2 while the average speed, $\bar{\lambda}' = 0.04$, is significantly lower than the standard adiabatic LZ process. This violates the traditional adiabatic theorem, which predicts better adiabaticity at lower speeds. Mapping the state evolution onto a Bloch sphere, the adiabatic process moves the state from the south pole to the north pole, while the modified process keeps the state near the south pole during the variation of the modified process, as shown in Fig. 1c.

To understand the counter-intuitive violation of adiabaticity in the above LZ tunneling process, we first examine the traditional adiabaticity criterion: $|\langle n(t)|m'(t)\rangle/(E_n(t) - E_m(t))| \ll 1$. Here, $|n(t)\rangle$ and $E_n(t)$ are the instantaneous eigenfunctions and eigenvalues satisfying $H(t)|n(t)\rangle = E_n(t)|n(t)\rangle$. This criterion suggests that adiabaticity improves with a smaller cross-Berry connection $A_{nm} = i\langle n(t)|m'(t)\rangle$ or a larger energy gap. Consequently, slower evolution is expected to enhance adiabaticity. However, about two decades ago, studies showed that this criterion fails in resonant phenomena, such as photoelectric effect and Rabi oscillation [3-8]. In the present case, no resonance occurs. To explain the adiabaticity violation, we introduce the instantaneous transition probability (ITP), and its time integral, instantaneous transition accumulation (ITA), to reveal the underlying physics.

**Instantaneous transition and its manipulation**

To understand the unexpected adiabaticity violation, we re-visit the adiabaticity condition for a time-dependent system governed by the Schrödinger equation $i\hbar|\Phi'(t)\rangle = H(t)|\Phi(t)\rangle$, with $\hbar = 1$ for simplicity. The instantaneous eigenenergies and eigenstates satisfy $H(t)|n(t)\rangle = E_n(t)|n(t)\rangle$, and the evolved state is $|\Phi(t)\rangle = \sum_n c_n(t)|n(t)\rangle$. The evolution is adiabatic if $|c_n(t_f)|^2 \approx |c_n(t_i)|^2$ for all eigenstates, where $t_i$ and $t_f$ are the initial and final times, respectively. After a small time interval $\delta t$, the amplitude of $n^{\text{th}}$ eigenstate is $c_n(t + \delta t) = \langle n(t + \delta t)|U(t + \delta t, t)|\Phi(t)\rangle$. Expanding $|c_n(t + \delta t)|^2$ to the first order in a Taylor series around $t$, we obtain

$$d|c_n(t)|^2/dt = Re(2i\sum_{m \neq n} c_m(t)c_n^*(t)A_{nm}(t)) \equiv Re(P_n(t)), \tag{1}$$

where $P_n(t) = 2i\sum_{m \neq n} c_m(t)c_n^*(t)A_{nm}(t)$ is the ITP for the $n^{\text{th}}$ band, $A_{nm}(t) = i\langle n(t)|m'(t)\rangle$ is the cross-Berry connection. Although Berry connection is gauge dependent, the combination with the coefficients $c$'s renders $P_n$ gauge independent [for more details, please see supplementary information (SI), section I]. ITA, the integral of ITP along the entire evolution process, whose real part $Re(A_n) = Re\left(\int_{t_i}^{t_f} P_n(t)dt\right)$, quantifies the population deviation of the $n^{\text{th}}$ band. Adiabaticity holds when $Re(A_n) \approx 0$ for all bands. Eq. (1) shows that the transition between bands depend on the cross-Berry connection and the coefficients of different bands, reflecting both the geometric and dynamic properties of the system. It should be noted that the traditional adiabaticity criterion can be derived from Eq. (1) under specific assumptions (SI, section II).

We next demonstrate that the temporal control over the evolution rate can serve as a tool to control the ITA and thus the population transfer. Because the real part of ITP governs the transition probabilities and directions, controlling the phase of $P_n$ is critical. This phase depends on the phase difference between two considered bands $c_m(t)c_n^*(t)$, and the corresponding cross-Berry connection $A_{nm}(t)$. For a system starting with nonzero population in both $n^{\text{th}}$ and $m^{\text{th}}$ bands, accumulating population in the $n^{\text{th}}$ band requires $P_n$ to remain on the non-negative half plane, as indicated in Fig. 2a. However, since the term $c_m(t)c_n^*(t)$ rotates on the complex plane, $Re(P_n)$ may become negative. To prevent this, we propose pausing the time variation of the Hamiltonian when $P_n$ approaches the negative half plane. During this pause, the cross-Berry connection becomes zero, setting $P_n = 0$, and halting population loss from the $n^{\text{th}}$ band. Meanwhile, $c_m(t)c_n^*(t)$ continues to rotate. After a half cycle, defined by the circular frequency $\Delta\omega = \omega_n -$

$\omega_m$, the Hamiltonian's time variation resumes, ensuring $Re(P_n)$ remains non-negative and population accumulates ($Re(A_n) > 0$). This approach, termed the phase difference manipulation (PDM) method, forms the core of our control strategy.

**Demonstration of Counter adiabatic LZ process**

Using the PDM method, we design a spoof surface plasmon polariton (sSPP) waveguide system (Fig. 2b) to directly observe the adiabaticity violation in modified LZ process. In this system, time $t$ is replaced by propagation distance $z$, and energy $E$ by propagation constant $\beta$. The Hamiltonian of the system can be expressed as $H_{LZ}(z) = \beta_0 I_2 + \kappa \sigma_x + \delta \sigma_z$, where $I_2$ is 2-by-2 identity matrix, $\kappa = \Delta\beta \sin(\lambda(z))$, $\delta = -\Delta\beta \cos(\lambda(z))$, $\Delta\beta = 0.0713 mm^{-1}$, and the bases are the eigenmodes of the two uncoupled waveguides. The onsite energies are tuned by the height $h$ of the tooth and the coupling is tuned by the distance $d$ between the two waveguides.

For a control study with a typical LZ process with uniform evolution speed, we set $\lambda(z) = \pi z/(201p)$, where $p = 1.33 mm$ is the period of the sSPP waveguide. And the onsite energy $\delta$ and coupling $\kappa$ are modulated according to Fig. 2c. The initial state is $|\Phi_0\rangle = |1\rangle = (0,1)^T = |\phi_+\rangle$. Although this process is nearly adiabatic, the populations of $|\phi_\pm(\lambda)\rangle$ exhibit slight variations and oscillations, as depicted by the red curves in Fig. 2d, e. These oscillations result from the back-and-forth transition driven by the ITP, $P_{-+} = 2ic_+(z)c_-^*(z)A_{-+}(z)$, from the upper to the lower energy level. Specifically, the lower band population $|c_-|^2$ decreases when $Re(P_{-+}) < 0$ and increases when $Re(P_{-+}) > 0$, as shown in Fig. 2d. Similarly, $|c_+|^2$ follows $Re(P_{+-})$, as shown in Fig. 2e, with $Re(P_{+-}) = -Re(P_{-+})$, ensuring conservation of the overall population.

To induce adiabaticity violation, we pause the Hamiltonian's evolution at specific points, disrupting the back-and-forth transitions. The results of this counter adiabatic LZ process are shown in Fig. 2f-h, with the evolution of $\delta$ and $\kappa$ depicted in Fig. 2f. Because $Re(P_{-+}) \geq 0$ for all $z$, the population of the lower level, $|c_-|^2$, continues to increase, as shown in Fig. 2g. Conversely, since $Re(P_{+-}) \leq 0$ for all $z$, the population of the upper level $|c_+|^2$ steadily decreases, as depicted in Fig. 2h. Thus, the adiabaticity is significantly violated despite a slower average evolution speed compared to the standard LZ evolution.

Using the effective parameters described above, we design the geometric parameters of the sSPP waveguide system under the paraxial approximation (More details of the sSPP waveguides are available in SI, section III). Three samples, operating at 24.4GHz, are designed and fabricated by a standard printed circuit board (PCB) method. The first two samples correspond to the effective parameters of the adiabatic and the counter-adiabatic processes shown in Fig. 2c and f, respectively. The third sample is a stretched version of the adiabatic sample, designed to match the evolution length of the counter-adiabatic sample.

In the measurement, an electromagnetic wave is injected into the 2$^{nd}$ waveguide, setting the initial state as $|\Phi(t_i)\rangle = |1\rangle = (0,1)^T = |\phi_+(t_i)\rangle$. For the adiabatic sample, both simulation (Fig. 3a) and measurement (Fig. 3b) show that the mode's population gradually transfers to the 1$^{st}$ waveguide, resulting in a final state $|\Phi(t_f)\rangle = |0\rangle = (1,0)^T = |\phi_+(t_f)\rangle$. Although the eigenmode's profile evolves, the index of the energy band remains unchanged, confirming adiabaticity. This evolution corresponds to the blue arrowed line shown in Fig. 1a. The adiabatic evolution of the longer adiabatic sample, shown in Fig. 3e, f, exhibits similar characteristics to the shorter adiabatic sample (More results are given in SI, section IV).

In contrast, for the counter adiabatic sample, both simulation (Fig. 3c) and measurement (Fig. 3d) reveal mode oscillations between the two waveguides. Oscillations observed arise because any state, including the initial state $|1\rangle = (0,1)^T$, tends to persist despite temporal changes in Hamiltonian. However, as the system's eigenstates evolve, the state $|\Phi(t)\rangle$ can be decomposed into the instantaneous eigenstates as $|\Phi(t)\rangle = |\Phi(t_i)\rangle = |1\rangle = (0,1)^T = c_+|\phi_+(t)\rangle + c_-|\phi_-(t)\rangle$. The difference in eigen-frequencies between $|\phi_+\rangle$ and $|\phi_-\rangle$ states causes beating, leading to population oscillation between the two waveguides, as both eigenstates have non-zero components in each waveguide. Despite multiple population oscillations between the two waveguides during the pauses, the mode ultimately resides in the 2$^{nd}$ waveguide, leading to a final state given by $|\Phi(t_f)\rangle = |1\rangle = (0,1)^T = |\phi_-(t_f)\rangle$. This switch from + to − energy level index confirms the violation of adiabaticity.

**Discussion and conclusion**

The above example demonstrates that the proposed instantaneous transition accumulation (ITA) provides a fundamental description of the adiabaticity of a parameter-varying process. Furthermore, the ITP serves as a controllable parameter for manipulating adiabaticity. Beyond this example, we explore additional intriguing cases in Sections V and VI of the SI, including initial-state-induced adiabatic evolution and shortcut-to-adiabaticity without altering the evolution path. These cases highlight that the PDM method is just one approach to utilizing ITP. Since ITP encapsulates the complex amplitudes of eigenstates, shaped by both geometric and dynamic phases, the interplay of these phases offers a promising avenue for controlling the evolution of quantum systems.

In summary, we have revisited the quantum adiabatic theorem and introduced ITA and ITP as fundamental physical quantities for characterizing adiabaticity in parameter-varying processes. By controlling ITA and ITP, we have developed phase difference manipulation (PDM), a strategy that enables precise control over adiabaticity in fixed-path parameter-varying processes. Using PDM, we have designed and experimentally demonstrated a photonic system exhibiting counter-intuitive violation of adiabaticity in a sufficiently slow Landau-Zener (LZ) process. Additionally, since Berry phase may emerge from adiabatic evolution, our findings suggest that even slow variations in an evolution process may affect the topological properties of a physical system, underscoring the need for closer scrutiny of the robustness of topological systems.

**Methods**

**Theoretical and numerical calculations:** By keeping the lowest non-zero time derivative, we get the ITP. The details of the derivation are given in SI, Section I. We use paraxial approximation to design the sSPP waveguide systems. The effective parameters of the Hamiltonians with respect to sSPP waveguide systems are extracted from full-wave simulation carried out by CST Studio. Then we calculate and design the evolutions of the effective Hamiltonians by discretized time ordering integral through MATLAB code. According to the designed effective Hamiltonian, we get the practical geometric parameters of the sSPP waveguide systems. And then the full-wave simulations are carried out for the entire sSPP waveguide systems via CST Studio to give the numerical evolution results.

**Sample preparation and experimental measurements:** The samples are fabricated using printed circuit board (PCB) technology and characterized using a microwave vector network analyzer (VNA). The spoof surface plasmon polariton (sSPP) mode is excited via a sub-miniature version A (SMA) port connected to the VNA, while its propagating field distribution in the waveguide is mapped using another probe antenna. Data acquisition is performed on a laptop interfaced with the VNA, and the probe's position is precisely controlled by a three-axis translational stage connected to the same computer. More experimental results are available in SI.

## Acknowledgments

This work was supported by the New Cornerstone Science Foundation and the Research Grants Council of Hong Kong (STG3/E-704/23-N, AoE/P-701/20, AoE/P-502/20, 17309021).

## Author contributions

O.Y. and S.Z. conceived this project. O.Y. derived the theoretical results. O.Y. and Z.J. performed analytical calculations and numerical simulations. O.Y. and Z.J. designed the samples. Z.J. performed the experiment. O.Y., Z.J., J. S., C. G., Q. D. and S. Z. participated in the analysis of the results. O.Y., Z.J. and S.Z. wrote the manuscript with input from all authors. C. G. and S. Z. supervised the project. All the authors contributed to discuss and review the manuscript.

## Competing interests

The authors declare no competing interests.

## Data and materials availability

All related data, code, and materials in the main text or the supplementary information are available by contacting corresponding authors.

## Supplementary information

Supplementary Information is available for this paper.

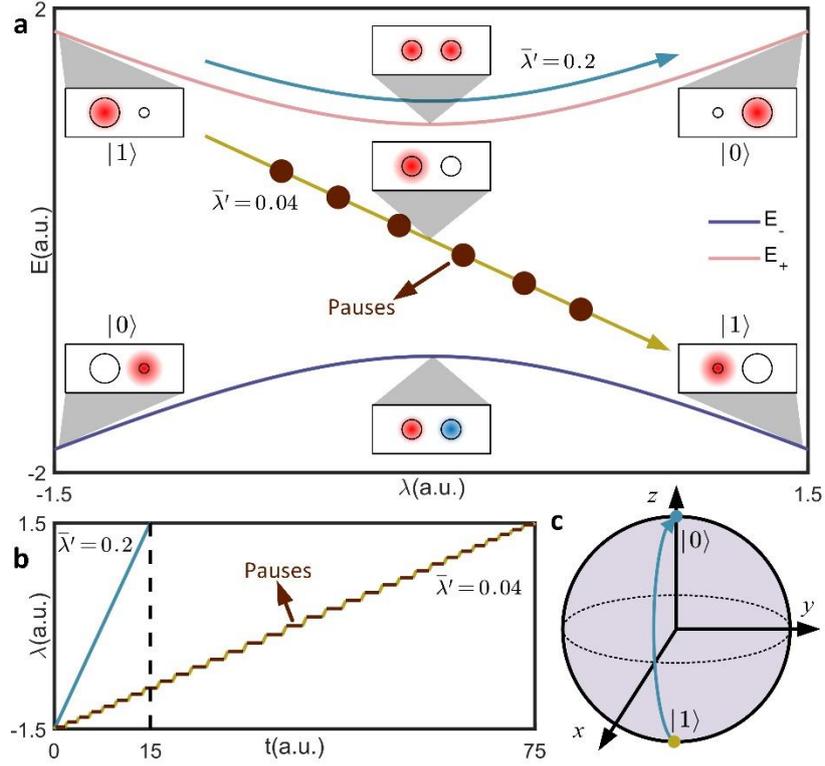

**Fig. 1 Adiabaticity violation in LZ process. a**, Schematic of adiabatic LZ process with average evolution speed $\bar{\lambda}' = 0.2$ which is indicated by the blue arrow line, and counter adiabatic LZ tunneling with $\bar{\lambda}' = 0.04$ which is indicated by the yellow arrow line. The counter adiabatic case stops evolution at some $\lambda$ points denoted by the red points. And its maximum $\lambda'$ equals 0.2, same as the adiabatic case. **b**, The time function of parameter $\lambda$ for the adiabatic LZ process is shown by the blue line, and the counter adiabatic case is shown by the yellow and red broken line. The yellow segments mean that the system is changing, and the red ones mean that the system pauses. The total evolution time of the counter adiabatic case is 5 times larger than the adiabatic one. **c**, The initial state $|1\rangle$ evolves along the blue path in adiabatic process and stays at $|1\rangle$ in counter adiabatic process when the system is changing.

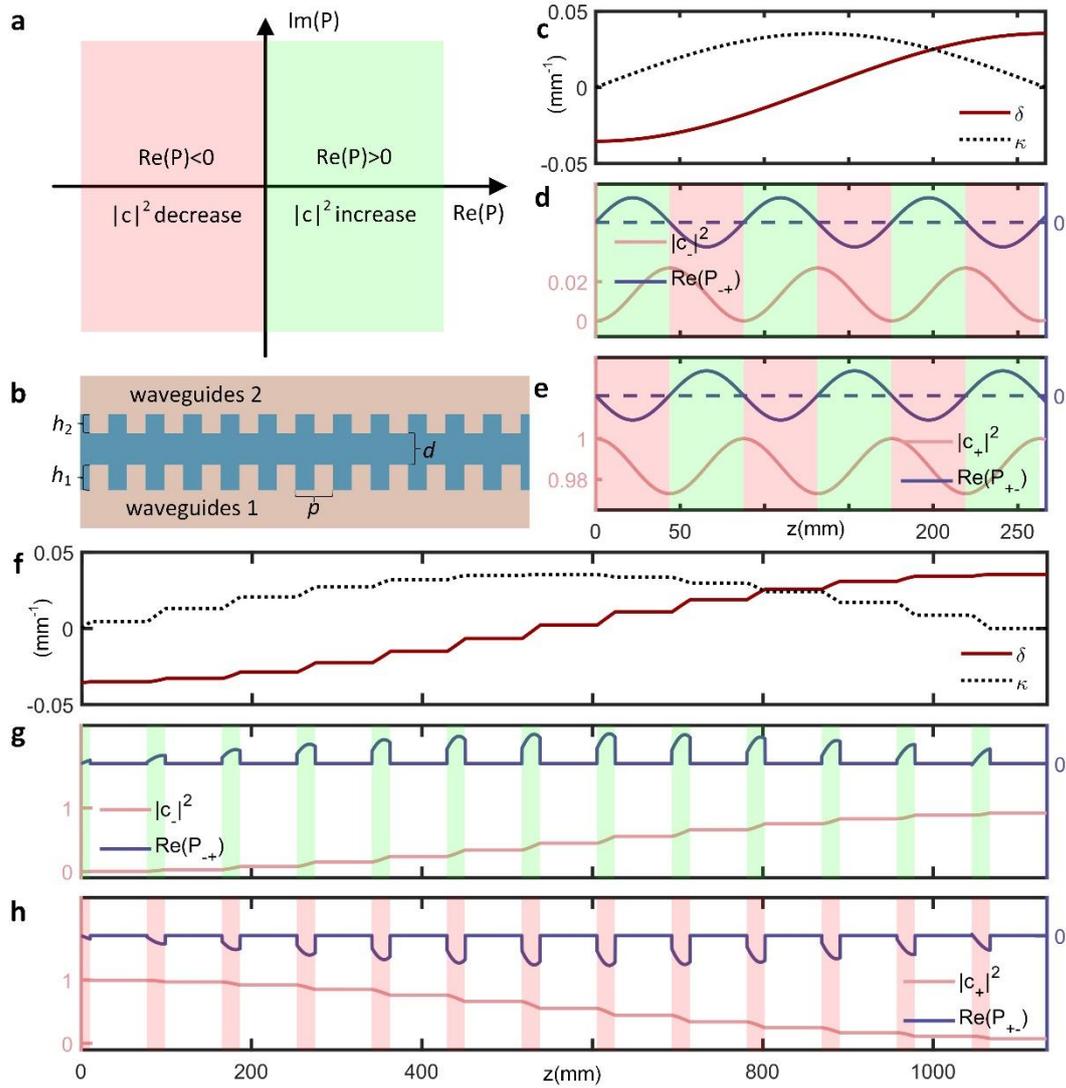

**Fig. 2 sSPP waveguide system designed by PDM to show adiabaticity violation. a**. A schematic showing how $Re(P)$ controls the population changing direction of an instantaneous eigenstate. **b**. The main geometric parameters of the sSPP waveguide system. **c**. Changing of onsite energy $\delta$ and coupling $\kappa$ of the sSPP waveguide system demonstrating an adiabatic LZ process. **d**. Back-and-forth transition of $|c_-|^2$ of the adiabatic LZ process determined by $Re(P_{-+})$. $|c_-|^2$ goes down when $Re(P_{-+}) < 0$, covered by red transparent rectangular, and goes up when $Re(P_{-+}) > 0$, covered by green transparent rectangular. **e**. Back-and-forth transition of $|c_+|^2$ of adiabatic LZ tunneling determined by $Re(P_{+-})$. **f**. Same as **c** but for counter adiabatic LZ process. **g**. $|c_-|^2$ keeps increasing as $Re(P_{-+}) \geq 0$ when the system evolves. **h**. $|c_+|^2$ keeps decreasing as $Re(P_{+-}) \leq 0$ when the system evolves. Note, the flat zero value segments of $Re(P)$ in **g** and **h** are caused by zero cross-Berry connection.

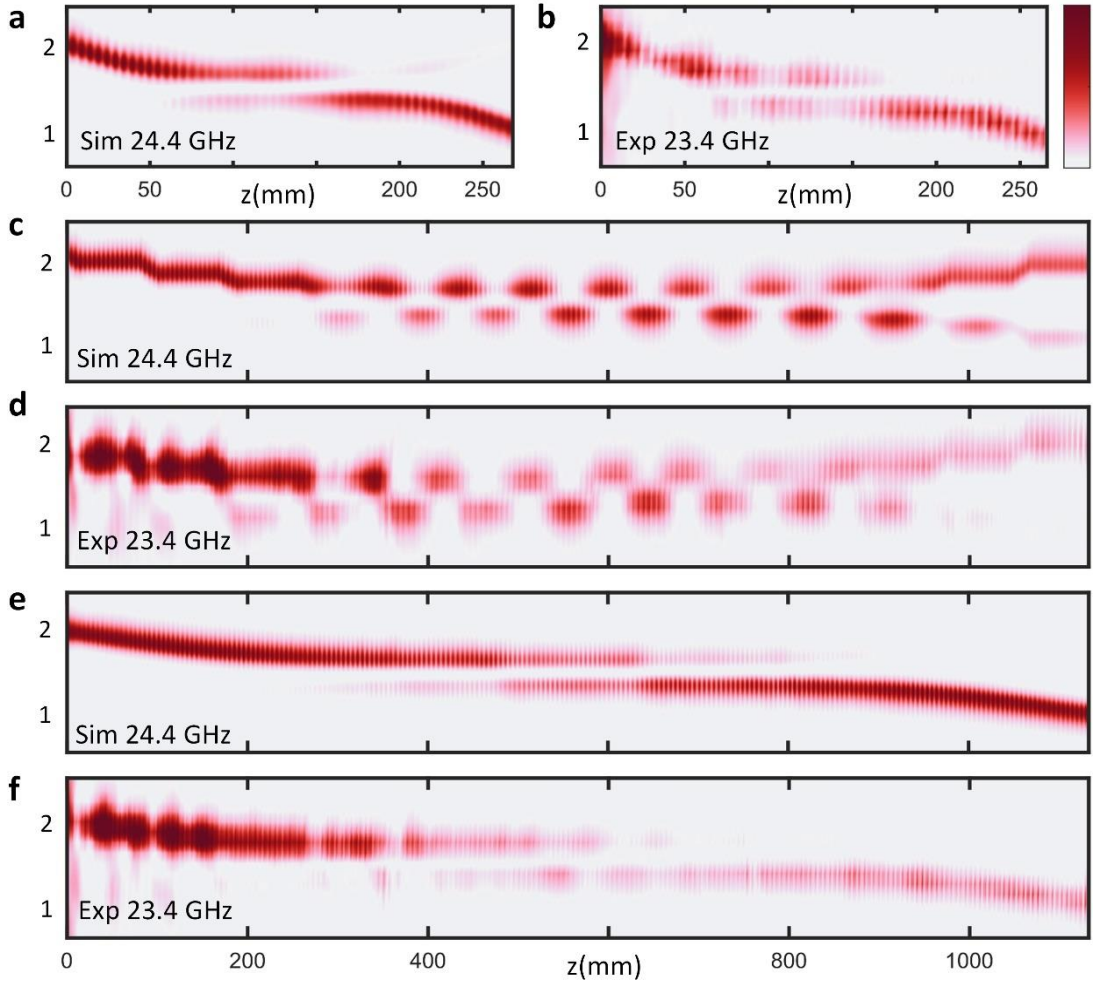

**Fig. 3 Simulated and experimental results. a.** Simulated result for adiabatic sSPP waveguide system at 24.4GHz. Light is injected into the second waveguide representing the initial state $|1\rangle = (0,1)^T$ and comes out from the first one representing the final state $|0\rangle = (1,0)^T$. This result matches well with the adiabatic evolution process shown in **Fig. 1** and **2**. **b.** Experimental result corresponding to **a**'s result with a slight frequency deviation due to fabrication error. **c.** Simulated result for counter adiabatic waveguide system whose variational Hamiltonian's parameters are shown in **Fig. 1f**. **d.** Same as **b** but for counter adiabatic evolution. **e, f.** Simulated and experimental results for long adiabatic sSPP waveguide system sharing the same evolution length with the counter adiabatic case. It's merely a stretched version of the adiabatic sample.